\documentclass[referee]{chjaa}

\usepackage{graphicx}
\input{epsf.sty}                        
\input{psfig.sty}                       

\begin{document}
\title{Hyperaccretion after the Blandford-Znajek Process:
a New Model for GRBs with X-Ray Flares Observed in Early Afterglows}

\volnopage{Vol.0 (200x) No.0, 000--000}
 \setcounter{page}{1}

\author{Wei-Hua Lei\mailto{}\ ,Ding-Xiong Wang,Yuan-Chuan Zou
\and Lei Zhang}
\institute{Department of Physics, Huazhong
University of Science and Technology, Wuhan, 430074, China
\email{leiwh@hust.edu.cn} }

\date{Received~~2007 month day; accepted~~2007~~month day}


\abstract{ We propose a three-stage model with Blandford-Znajek
(BZ) and hyperaccretion process to interpret the recent
observations of early afterglows of Gamma-Ray Bursts (GRBs). In
the first stage, the prompt GRB is powered by a rotating black
hole (BH) invoking the BZ process. The second stage is a quiet
stage, in which the BZ process is shut off, and the accretion onto
the BH is depressed by the torque exerted by the magnetic coupling
(MC) process. Part of the rotational energy transported by the MC
process from the BH is stored in the disk as magnetic energy. In
the third stage, the MC process is shut off when the magnetic
energy in the disk accumulates and triggers the magnetic
instability. At this moment, the hyperaccretion process may onset,
and the jet launched in this restarted central engine generates
the observed X-ray flares. This model can account for energies and
timescales of GRBs with X-ray flares observed in early afterglows.
}

\authorrunning{W.-H. Lei, D.-X. Wang, Y.-C. Zou \& L. Zhang}
\titlerunning{A new model for GRBs with X-ray flares}  

\maketitle

\section{INTRODUCTION}
\label{sect:intro} Recently, the flares in the early X-ray
afterglows have been discovered in both long and short bursts with
the X-ray telescope (XRT) on board Swift. These flares appear at
about tens to hundreds seconds after the trigger of the GRBs and
last about several hundred seconds (for review of the Swift results
and, in paricular, the X-ray flares, see Zhang 2007). Their rapid variability has been
interpreted as the inner engine being active much longer than the
duration of the GRB itself (Zhang et al. 2006). Therefore, it is
needed that after the cease of the prompt gamma-ray emission, the
central engine can be restarted (Fan {\&} Wei 2005; Zhang et al.
2006).

For the case of long GRBs, King et al. (2005) suggested that the
X-ray flares could be produced from the fragmentation of the
collapsing stellar core in a modified hypernova scenario. The
fragment subsequently merges with the main compact object formed in
the collapse, releasing extra energy. In this two-stage collapse
model, the time delay between the burst and the flare reflects the
gravitational radiation time scale for the orbiting fragment to be
dragged in.

For the case of short GRBs, MacFadyen, Ramirez-Ruiz {\&} Zhang
(2005) suggested that the flares could be the result of the
interaction between the GRB outflow and a non-compact stellar
companion, in this model, short GRBs result from the collapse of a
rapidly rotating neutron star in a close binary system. While Dai
et al. (2006) suggested that the flares in short GRBs can be
produced by differentially rotating, millisecond massive pulsars
after the mergers of binary neutron star. The differential
rotation leads to windup of interior poloidal magnetic fields and
the resulting toroidal fields are strong to float up and break
through the stellar surface. Magnetic reconnection-driven
explosive events then occur, leading to multiple X-ray flares
minutes after the original gamma-ray burst. Gao {\&} Fan (2006)
also suggested a short-lived supermassive magnetar model to
account for the X-ray flares following short GRBs. In their model,
the X-ray flares are powered by the dipole radiation of the
magnetar.

Motivated by the fact that the flares are observed in both long
and short duration GRBs, Perna et al. (2005) suggested that the
flares could be produced due to subsequent accretion of blobs of
material in the hyperaccreting accretion disk, which initially
circularize at various radii and subsequently evolve viscously.
While Proga {\&} Zhang (2006) conjectured that the energy release
can be repeatedly stopped and then restarted by the magnetic flux
accumulated around the accretor. The restarting of the~accretion
corresponds to the X-ray flares.

Although the main features of the X-ray flares can be interpreted
successfully in the above scenarios, a quantitative model with a
clear physical process for restarting the central engine remains
lacking.

Very recently, Lei et al. (2005a, hereafter L05; 2005b) proposed a
scenario for GRBs in Type Ib/c SNe, invoking the coexistence of
the Blandford-Znajek (BZ; Blandford {\&} Znajek 1977) and Magnetic
Coupling (MC; Blandford 1999; van Putten 1999; Li 2000, 2002; Wang
et al. 2002) processes. The BZ process can provide ``clean''
energy for the prompt GRB by extracting rotating energy from a BH
(Lee et al. 2000). In L05, the accretion onto the BH is depressed
by the torque exerted by the BH through MC process. It is found
that the BZ process will be shut off before the MC process because
of the close of the half-opening angle on the BH horizon.
Recently, van Putten {\&} Levinson (2003, hereafter PL03)
suggested that the magnetic energy accumulated by the MC process
can trigger the magnetic instability, which results in the cease
of the MC process (Eikenberry \& van Putten 2003).

Based on L05 and PL03, we suggest that the pause and the restart of
the central engine arise from the MC process, and propose a
three-stage model to interpret the GRBs and their flares. In this
model, the prompt GRB and the X-ray flare are powered by the BZ
process and the subsequent hyperaccretion, respectively.

\section{THE THREE-STAGE MODEL}

In this section, we present the three-stage model for the inner
engine for GRBs. The magnetic field configuration is shown in Figure
1, which is adapted from van Putten (2001).

\begin{figure}
\begin{center}
\mbox{\epsfxsize=0.5\textwidth\epsfysize=0.5\textwidth\epsfbox{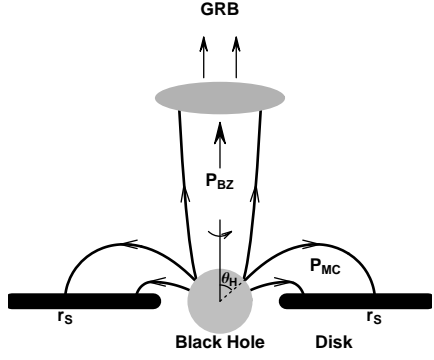}}
\caption{A schematic drawing of the magnetic field configuration of
a three-stage model}
\end{center}
\end{figure}

In Figure 1 the angle $\theta _H $ is the half-opening angle of the
open magnetic flux tube, indicating the angular boundary between
open and closed field lines on the horizon. The angle $\theta _H $
can be determined by (Wang et al. 2003)

\begin{equation}
\label{eq1} \cos \theta _H = \int_1^{\xi _S } {\mbox{G}\left( {a_ *
;\xi ,n} \right)d\xi }.
\end{equation}
where, $a_\ast \equiv J / M^2$ is the BH spin defined in terms of
the BH mass $M$ and the angular momentum $J$, the parameter $n$ is
the power-law index for the variation of $B_{D}^P$,
i.e.,$B_{D}^P\propto \xi ^{ - n}$,and $\xi \equiv r / r_{ms} $ is
the radial coordinate on the disk, which is defined in terms of
the radius $r_{ms} \equiv M\chi _{ms}^2 $ of the marginally stable
orbit (Novikov {\&} Thorne 1973). $\chi _{ms}$ depends only on
$a_*$ as (Novikov {\&} Thorne 1973)
\begin{equation}
\chi_{ms}^4- 6 \chi_{ms}^2+8 a_* \chi_{ms}-3a_*^2=0.
\end{equation}
The function $\mbox{G}\left( {a_ * ;\xi ,n} \right)$ is given by

\begin{equation}
\label{eq2} \mbox{G}\left( {a_ * ;\xi ,n} \right) = \frac{\xi ^{1 -
n}\chi _{ms}^2 \sqrt {1 + a_ * ^2 \chi _{ms}^{ - 4} \xi ^{ - 2} +
2a_ * ^2 \chi _{ms}^{ - 6} \xi ^{ - 3}} }{2\sqrt {\left( {1 + a_
* ^2 \chi _{ms}^{ - 4} + 2a_ * ^2 \chi _{ms}^{ - 6} }
\right)\left( {1 - 2\chi _{ms}^{ - 2} \xi ^{ - 1} + a_ * ^2 \chi
_{ms}^{ - 4} \xi ^{ - 2}} \right)} }.
\end{equation}

The parameter $\xi_{S}\equiv r_{S}/r_{ms}$ in the equation
(\ref{eq1}),is the critical radius of screw instability in the MC
process, which is determined by the criteria for the screw
instability given by Wang et al. (2004), i.e.,

\begin{equation}
\label{eq3} \frac{B_D^P}{B_D^T}<{\frac{L}{2\pi\varpi_D}},
\end {equation}
where $L$ is the poloidal length of the closed field line, and
$\varpi_D$ is the cylindrical radius on the disk, $B_D^P$ and
$B_D^T$ are the poloidal and toroidal componets of the magnetic
field on the disk, respectively. $2 \pi \varpi_D/L$ and
$B_D^P/B_D^T$ are expressed as (Wang et al. 2004)
\begin{equation}
B_D^P/B_D^T = \frac{\xi ^{1 - n}\left( {1 + q} \right)\left[
{2\csc ^2\theta - \left( {1 - q} \right)} \right]}{2a_ * \left( {1
- \beta } \right)}\sqrt {\frac{1 + a_ * ^2 \chi _{ms}^{ - 4} \xi
^{ - 2} + 2a_ * ^2 \chi _{ms}^{ - 6} \xi ^{ - 3}}{1 + a_ * ^2 \chi
_{ms}^{ - 4} + 2a_ * ^2 \chi _{ms}^{ - 6} }} ,
\end{equation}

\begin{equation}
\frac{L}{2 \pi \varpi_D} = \frac{1}{4\xi \sqrt{1+a_*^2 \xi^{-2}
\chi_{ms}^{-4}+ 2 a_*^2 \xi^{-3} \chi_{ms}^{-6}}} \int_{\xi_H}
^\xi \frac{d\xi}{\sqrt{1+a_*^2 \xi^{-2} \chi_{ms}^{-4}-2\xi^{-1}
\chi_{ms}^{-2}}}.
\end{equation}

By using equation (\ref{eq1}) we have the critical lines for the
half-opening angle $\theta_H =0$ in $a_\ast-n$ parameter space as
shown in Figure 2.

\begin{figure}
\begin{center}
\mbox{\epsfxsize=0.5\textwidth\epsfysize=0.5\textwidth\epsfbox{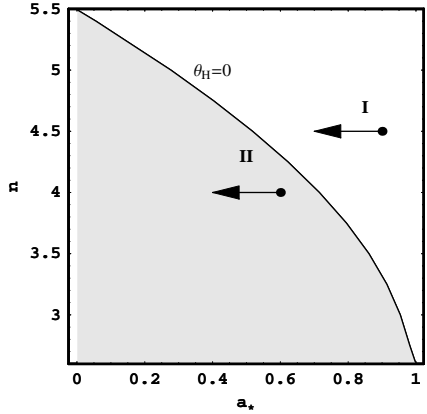}}
\caption{ The critical lines for screw instability in   parameter
space, by which the space is divided into regions \textbf{I} and
\textbf{II}. Each filled circle with arrowhead in the regions
represents one evolutionary state of the BH.}
\end{center}
\end{figure}

An interesting feature shown in Figure 2 is that the angle
$\theta_H$ can evolve to zero (the shaded region) with the
decreasing $a_\ast$ for $2.616\le n \le 5.493$. This evolution
characteristic implies that the BZ process will be shut off when
the BH spin decreases to the critical value $a_\ast^{GRB}$
corresponding to $\theta_H=0$. The lifetime of the half-opening
angle $\theta_H$ is defined as the evolution time of BH from the
initial spin $a_\ast(0)$ to $a_\ast^{GRB}$.

\subsection{Stage 1: the Prompt Emission of GRB}

In the first stage, the BZ process works together with the MC
process and the evolution state of the BH is located in region
\textbf{I} of Figure 2.

Since angular momentum is transferred from the rapidly rotating BH
to the disk, a very strong torque will be exerted on the inner
disk in the MC process. The accretion onto the BH will be
depressed by the MC process (van Putten {\&} Ostriker 2001, PL03;
Li 2002). Therefore, in this stage, the accretion rate is very
low, and the prompt emission of GRB is mainly powered by the BZ
process.

The powers and torques for the BZ and MC processes are expressed
as (Wang et al. 2002; Wang et al. 2003)

\begin{equation}
\label{eq4} \tilde {P}_{BZ} \equiv {P_{BZ} } \mathord{\left/
{\vphantom {{P_{BZ} } {P_0 }}} \right. \kern-\nulldelimiterspace}
{P_0 } = 2a_ * ^2 \int_0^{\theta _{H} } {\frac{k\left( {1 - k}
\right)\sin ^3\theta d\theta }{2 - \left( {1 - q} \right)\sin
^2\theta }} ,
\end{equation}

\begin{equation}
\label{eq5} {\tilde {T}_{BZ} \equiv T_{BZ} } \mathord{\left/
{\vphantom {{\tilde {T}_{BZ} \equiv T_{BZ} } {T_0 }}} \right.
\kern-\nulldelimiterspace} {T_0 } = 4a_ * \left( {1 + q}
\right)\int_0^{\theta _{H} } {\frac{\left( {1 - k} \right)\sin
^3\theta d\theta }{2 - \left( {1 - q} \right)\sin ^2\theta }} ,
\end{equation}

\begin{equation}
\label{eq6} {\tilde {P}_{MC} \equiv P_{MC} } \mathord{\left/
{\vphantom {{\tilde {P}_{MC} \equiv P_{MC} } {P_0 }}} \right.
\kern-\nulldelimiterspace} {P_0 } = 2a_ * ^2 \int_{\theta _{H}
}^{\pi / 2} {\frac{\beta \left( {1 - \beta } \right)\sin ^3\theta
d\theta }{2 - \left( {1 - q} \right)\sin ^2\theta }} ,
\end{equation}

\begin{equation}
\label{eq7} {\tilde {T}_{MC} \equiv T_{MC} } \mathord{\left/
{\vphantom {{\tilde {T}_{MC} \equiv T_{MC} } {T_0 }}} \right.
\kern-\nulldelimiterspace} {T_0 } = 4a_ * \left( {1 + q}
\right)\int_{\theta _{H} }^{\pi / 2} {\frac{\left( {1 - \beta }
\right)\sin ^3\theta d\theta }{2 - \left( {1 - q} \right)\sin
^2\theta }} ,
\end{equation}

\noindent where we have $q \equiv \sqrt {1 - a_\ast ^2 } $, $P_0
\equiv B_H^2M^2$ and $T_0\equiv B_H^2M^3$. $B_H $ is the magnetic
field on the BH horizon. The parameters $k$ and $\beta $ are the
ratios of the angular velocities of the open and closed magnetic
field lines to that of the BH, respectively. Usually, $k = 0.5$ is
taken for the optimal BZ power.

The energy $E_{BZ}$ extracted in the BZ process can be calculated
during the evolution of GRB. The true energy for GRBs, $E_\gamma$,
is given in L05,

\begin{equation}
\label{eq8} E_\gamma = \varepsilon_\gamma E_{BZ} =2.69\times
10^{53}ergs \times \left(\frac{M(0)}{M_{\sun}}\right) \left(
\frac{\varepsilon_\gamma}{0.15}\right)\int_{a_\ast (0)}^{a_\ast
^{GRB} }{\frac{\tilde{M}
\tilde{P}_{BZ}}{-\tilde{T}_{mag}+2a_\ast\tilde{P}_{mag}} }da_\ast,
\end{equation}
where $\tilde{P}_{mag} \equiv \tilde{P}_{BZ}+\tilde{P}_{MC}$,
$\tilde{T}_{mag} \equiv \tilde{T}_{BZ}+\tilde{T}_{MC}$, and
$\tilde{M}\equiv M/M(0)$ is the ratio of the BH mass to its initial
value. Following van Putten et al. (2004), we take the efficiency
$\varepsilon_\gamma=0.15$ in calculations.

The duration of GRB, $t_{GRB}$, is defined as the lifetime of the
half-opening angle $\theta_H$, which is exactly equal to the time
for the BH spin evolving from $a_\ast(0)$ to $a_\ast^{GRB}$, i.e.,

\begin{equation}
\label{eq9} t_{GRB}= 2.7\times10^3s\times \left( \frac{10^{15}G}{B_H
} \right)^2\left(\frac{M_{\sun}}{M(0)}\right)\int_{a_\ast
(0)}^{a_\ast ^{GRB}
}{\frac{\tilde{M}^{-1}}{-\tilde{T}_{mag}+2a_\ast\tilde{P}_{mag}}
}da_\ast,
\end{equation}

The energy and duration of the prompt emission can be obtained by
taking $M(0)$, $a_\ast(0)$, $n$ and $B_H$ into equations (\ref{eq8})
and (\ref{eq9}). In the following calculations the initial BH mass
and spin are assumed to be $M(0)=7M_{\sun}$ and $a_\ast(0)=0.9$.

By adopting proper values of parameters $n$ and $B_H$, our model can
explain both long and short bursts, such as GRB 050219A and GRB
050709 (see Table 1).


\begin{table}[]
\caption[]{GRB 050219A and GRB 050709. }
\begin{minipage}{\textwidth}
\begin{center}
\begin{tabular}{cccccc}
\hline\noalign{\smallskip}

GRBs & $Type$  & $T_{90}(s)$  & $E_\gamma (10^{50}erg)$  & $n$ & $B_H(10^{15}G)$ \\
\hline\noalign{\smallskip}

50219A & $Long$  & 23.6\footnote{Tagliaferri et al. 2005}  &
$>4.8^a$
& $>$3.545  & $>$0.7 \\

050709 & $Short-hard$ & 0.07\footnote {Fox et al. 2005} & $0.021^b$ & 3.330  & 2 \\

\noalign{\smallskip}\hline
\end{tabular}
\end{center}
\end{minipage}
\end{table}

From Table 1 we find that the shorter duration and lesser energy of
short burst could be fitted with smaller value of $n$ and greater
value of $B_H$. Some long bursts have been fitted in L05 by
reasonable values of $n$ and $B_H$.

\subsection{Stage 2: Accretion Depressed by the MC Process }
In this stage, the MC process is still active, and it depresses
the accretion onto the BH. Following PL03, the magnetic field
energy of disk $\varepsilon_B$ will build up in response to the
power received from the BH, i.e.,

\begin{equation}
\label{eq10} \varepsilon_B(t) = \int_{0}^{t} {\eta P_{MC}}dt,
\end{equation}
where, $\eta$ is the ratio of the increase rate of the magnetic
field energy to the total MC power, and $\eta \ll 1$ since most of
the rotational energy of a Kerr BH is emitted in gravitational
radiation and MeV neutrino emissions (van Putten 2001).

According to PL03 the magnetic instability of the disk will occur if

\begin{equation}
\label{eq11} \varepsilon_B/\varepsilon_K >1/15,
\end{equation}
where $\varepsilon_K$ is the kinetic energy of disk. Once this
instability is triggered, the magnetic field lines connecting the BH
horizon and the disk will disconnect quickly, i.e., the MC process
will be ceased.

As the accretion in the inner part of the disk is blocked by the MC
process the inner region will pile up at the accretion rate
corresponding to the radius $r_S$, and a torus might form due to the
matter accumulation.

In the case without the depression of the MC process the accretion
rate can be estimated by considering the balance between the
pressure of the magnetic field and the ram pressure of the innermost
part of the accretion flow (Moderskin et al. 1997), i.e.,

\begin{equation}
\label{eq12} B_H^2/(8\pi) = P_{ram} \sim \rho c^2\sim
\dot{M}_D/(4\pi r_H^2),
\end{equation}

As seen in Figure 1, the MC torque mainly affects the accretion flow
within the radius $\xi_S$. As a simple model, we assume the
accretion rate of the disk beyond $\xi_S$ can also be estimated by
equation (\ref{eq12}). The time of forming torus in the suspended
accretion state can be estimated as

\begin{equation}
\label{eq13} \tau_{torus} \approx \frac{M_T}{\dot{M}_D}= \alpha
B_{15}^{-2} m_H^{-1}\frac{2}{\left({1+q}\right)^2}2.7\times 10^3s,
\end{equation}
where $M_T$ is the mass of torus, and $m_H\equiv M/M_{\sun}$,
$\alpha \equiv M_T/M$, $B_{15}\equiv B_H/{10^{15}G}$. For
$M_T=0.1M_{\sun}$, $B_H=10^{15}G$ and $a_\ast=0.5$, we have
$\tau_{torus} \approx 3.2s$. Thus a torus will form after (for a
short bursts) or during (for a long bursts) the prompt emission of
GRB. The kinetic energy of the torus $\varepsilon_K$ can be
estimated as (PL03)

\begin{equation}
\label{eq14} \varepsilon_K=\frac{M_TM}{2r_T},
\end{equation}
where, the radius of the torus $r_T$ can be estimated by $r_T\approx
r_{ms}$.

Once the magnetic field energy of disk $\varepsilon_B$ satisfies
equation (\ref{eq11}), the instability occurs, and the magnetic
field lines connecting the BH horizon and the disk will be
disconnected quickly. At this time, the MC process ceases, and the
BH spin is denoted as $a_*^{acc}$.

\subsection{Hyperaccretion as the Restarting Engine for X-Ray
Flares} The torque to depress the accretion flow will disappear
immediately once the MC process stops. At this stage the viscous
torque is dominative, and the hyperaccretion will occurs due to an
abrupt lose of angular moment of disk matter.

Since the BZ process has been ceased at the end of the first stage
due to the shutting off of the half-opening angle $\theta_H$, the
only operating energy process at this time (the third stage) is
hyperaccretion.

We assume that the X-ray flares are powered by the hyperaccretion,
of which the starting time is denoted as $t_{acc}$ (in the frame of
central engine). In addition, we assume that the starting time
$t_{acc}$ is determined by the duration for the BH spin evolving
from $a_\ast(0)$ to $a_\ast^{acc}$.

The observational beginning time of the flare is $t_{flare}$ .We
relate $t_{acc}$ to the radius for the new ejecta (produced by the
hyperaccretion process) to catch up with the foregoing merged
shell of GRB prompt stage by

\begin{equation}
\label{eq15} R_f \sim \gamma^2ct_{acc},
\end{equation}

Here the Lorentz factors of the ejected shells and their diversity
are all assumed in the same order $\gamma$(Sari {\&} Piran 1995).
Considering that the observational time obeys $t_{flare}\approx
R/\gamma^2c$, we can regard $t_{acc}$ as the time of observing the
X-ray flare, i.e., $t_{flare} \sim t_{acc}$.

In our model $t_{acc}$ is sensitive to the values of $n$, $B_H$ and
$\eta$, where $n$ and $B_H$ are obtained by fitting the energy and
duration of the GRB prompt emssion (See the discussion in stage 1).
Therefore, $\eta$ is only determined by $t_{acc}$ or $t_{flare}$ for
a given GRB. For GRB 050219A and GRB 050709, the X-ray flares are
observed at $t_{flare}\sim 100s$ after the trigger of the bursts.
The values of the parameter are taken as $\eta =0.05$ and 0.01 for
GRB 050219A and GRB 050709 in calculations, respectively.

For other bursts, the values of $\eta$ are in the same order. This
can be proved by the following estimation. For $M_T = 0.1M_{\sun}$,
the critical magnetic energy is
$\varepsilon_B=\varepsilon_K/15\approx 10^{51}erg$. The MC power can
be estimated as $P_{MC} \sim B^2M^2\approx 6.59\times 10^{50}erg/s$.
If $t_{flares}\sim 100s$, we have $\eta \sim 0.02$.

It has been argued that the hyperaccreting torus can power the X-ray
flares via neutrino annihilation (e.g. Ruffert {\&} Janka 1998). The
jet luminosity driven by this mechanism depends on the mass
accretion rate very sensitively, since the neutrino emission is
related very closely to the density and the temperature of the
torus. For accretion rates ($\tilde{M_D}$) between 0.01 and
$0.1M_es^{-1}$, the $\nu\bar{\nu}$ annihilation luminosity can be
well fitted by (Popham et al. 1999; Fryer et al. 1999; Janiuk et al.
2004)

\begin{equation}
\label{eq16} log L_{\nu\bar{\nu}}(erg \cdot s^{-1})\approx
43.6+4.89log\left(\frac{\dot{M}}{0.01M_{\sun}
s^{-1}}\right)+3.4a_\ast,
\end{equation}

\noindent In this stage, the BH spin must be $a_\ast^{acc}$, which
is about 0.5.

The total energy of the flare is also typically smaller than that of
the prompt emission, although in some cases both could be comparable
(e.g. for GRB 050502B). Based on equation (\ref{eq16}), a torus with
mass of $0.1M_{\sun}$ can provide $10^{50}erg$ via neutrino
annihilation if its accretion rate is about $0.1M_{\sun}/s$. This
energy is enough to power the X-ray flares in GRB afterglow.

The duration of the flare can be estimated as following. As
suggested by Zou, Dai {\&} Xu (2006), we assume that the emission of
X-ray flares come from the reverse shock of internal shock. For a
relativistic reverse shock, the duration of the shock in the
observer's frame is $\Delta t _{flare} \simeq \Delta /c$(Sari {\&}
Piran 1995), while the depth of shell for the spreading case is
$\Delta \simeq R/\gamma^2$. Thus, the duration of one flare is

\begin{equation}
\label{eq17} \Delta t_{flare} \simeq R_f/ \gamma^2c \simeq t_{flare}
\end{equation}

It is interesting to note that the three time scales are equal
approximately, $\Delta t_{flare} \sim t_{flare} \sim t_{acc}$, and
they are all independent of the Lorentz factors. This result is
also in agreement with the observations.

Usually, the afterglow light curves are characterized by multiple
flares. This feature could be understood based on restarting the
hyperaccretion several times, provided that the disk may be
fragmentized by the gravitational instability as proposed by Perna
et al. (2005), or the accretion can be repeatedly stopped and then
restarted by the magnetic flux accumulated as shown by Proga {\&}
Zhang (2006).

Moreover, the durations of these flares seem to be positively
correlated with the epochs when the flares happen, i.e., the later
the epoch, the longer the duration. There is particularly evident in
GRB 050502B and the short-hard GRB 050724. In both cases, there is
an early flare (several 100s for GRB 050502B and several 10s for GRB
050724) whose duration is of the order of the peak time itself, and
there is also a very late flare at tens of thousands seconds with a
duration of the same order. The above properties of the flares could
be interpreted by equation (\ref{eq17}), which predicts that the
arrival time of each new flare should directly correlate with its
total duration.

\section{Discussion}

In this paper, we propose a three-stage model for the inner engine
for GRBs. The prompt emission of GRB is explained by the BZ
process, and the X-ray flares are explained by the latter
hyperaccretion process. One of the features of our model is the MC
effects on the accretion process, both the depress accretion and
the start of the hyperaccretion process are related intimately to
the MC mechanism. It turns out that some features of the X-ray
flares in the early afterglow after prompt gamma ray emission can
be interpreted.

Comparing with the models of Perna et al. (2005) and Proga {\&}
Zhang (2006), our model can describe the cease and the restart of
the central engine in a quantitative way. While the former have
their advantages in explaining the multiple flares. Our model can
also interpret the multiple flares if the MC process can be stopped
and restarted repeatedly. Although the restart of the MC process is
very speculative, it is possible since the magnetic filed can be
recovered by the dynamo process (Hawley et al. 1995) or the
accumulation of the magnetic flux (Spruit, Stehle and Papaloizou
1995).

In our model, the X-ray flares are powered by the hyperaccretion
process. However, for short GRBs, Fan et al. (2005) suggested that
this process is not efficient. Instead, the BZ process is a very
powerful mechanism. Similar results are found in Xie et al. (2007).
Like the MC process, the restart of the BZ process is also possible.

From the above discussion we find that the rebuilding of the
magnetic field is essential to our model, and we intend to address
this issue in our future work.

\section{Acknowledgments}
We thank the anonymous referee for constructive suggestions. We
also thank Y. Z. Fan for helpful discussions. This work is
supported by the National Natural Science Foundation of China
under Grant Number 10703002.


\label{lastpage}

\end{document}